\documentclass[newfonts=false,format=sigconf,10pt,letterpaper,nonacm]{acmart}

\usepackage[utf8]{inputenc}
\usepackage{xcolor}
\usepackage{hyperref}
\usepackage{booktabs} 
\usepackage{subcaption}
\usepackage{caption}
\usepackage[inline]{enumitem} 

\usepackage[disable]{todonotes}

\newcommand{\IC}[1]{\todo[inline,color=green!20]{IC: #1}}
\newcommand{\YBI}[1]{\todo[inline,color=orange!20]{YBI: #1}}

\newcommand{\TRIANGLE}{triangle}
\newcommand{\TRIANGLEUC}{Triangle}
\newcommand{\BESTTRIANGLE}{best-triangle}
\newcommand{\BESTTRIANGLEUC}{Best-triangle}
\newcommand{\RELAY}{relay}
\newcommand{\RELAYUC}{Relay}
\newcommand{\BESTRELAY}{best-relay}
\newcommand{\BESTRELAYUC}{Best-relay}
\newcommand{\GRAPHITE}{Graphite}
\newcommand{\GRAFANA}{Grafana}
\newcommand{\CLOUDMANAGER}{Cloud Manager} 

\newcommand{\T}[1]{\noindent\textbf{#1}}
\newcommand{\TT}[1]{\textit{{#1}}}

\title{CloudCast: Characterizing Public Clouds Connectivity}
\date{January 2019}
\setcopyright{none}
\renewcommand\footnotetextcopyrightpermission[1]{} 
\newif\ifblind
\blindfalse

\begin{document}

\newcommand{\aut}[2]{#1\texorpdfstring{$^{#2}$}{(#2)}} 
\author{Noga H. Rotman}
  \affiliation{
    \institution{Hebrew University of Jerusalem}
    \country{}}
  \email{nogar02@cs.huji.ac.il}
\author{Yaniv Ben-Itzhak}
  \affiliation{
    \institution{VMware}
    \country{}}
  \email{ybenitzhak@vmware.com}
\author{Aran Bergman}
  \affiliation{
    \institution{VMware}
    \country{}}
  \email{bergmana@vmware.com}
\author{Israel Cidon}
  \affiliation{
    \institution{VMware}
    \country{}}
  \email{icidon@vmware.com}
\author{Igor Golikov}
  \affiliation{
    \institution{VMware}
    \country{}}
  \email{igolikov@vmware.com}
\author{Alex Markuze}
  \affiliation{
    \institution{VMware}
    \country{}}
  \email{amarkuze@vmware.com}
\author{Eyal Zohar}
  \affiliation{
    \institution{VMware}
    \country{}}
  \email{eyalzo@gmail.com}



\begin{abstract}
Public clouds are one of the most thriving technologies of past decade.
Major applications over public clouds require world-wide distribution and large amounts of data exchange between their distributed servers. 
To that end, major cloud providers have invested tens of billions of dollars in building world-wide inter-region networking  infrastructure that can support high performance communication into, out of, and across public cloud geographic regions.

In this paper, we lay the foundation for a comprehensive study and real time monitoring of various characteristic of networking within and between public clouds. We start by presenting CloudCast, a world-wide and expandable measurements and analysis system, currently (January 2019) collecting data from three major public clouds (AWS, GCP and Azure), 59 regions, 1184 intra-cloud and 2238 cross-cloud links (each link represents a direct connection between a pair of regions), amounting to a total of 3422 continuously monitored links and providing active measurements every minute. CloudCast is composed of measurement agents automatically installed in each public cloud region, centralized control, measurement data base, analysis engine and visualization tools.

Then we turn to analyze the latency measurement data collected over almost a year .
Our analysis yields surprising results. First, each public cloud exhibits a unique set of link latency behaviors along time. Second,
using a novel, fair evaluation methodology, termed \textit{similar links}, we compare the three clouds. Third, we prove that more than 50\% of all links do not provide the optimal RTT through the methodology of \textit{\TRIANGLE{}s}. \TRIANGLEUC{}s also provide a framework to get around bottlenecks, benefiting not only the majority (53\%-70\%) of the cross-cloud links by 30\% to 70\%, but also a significant portion (29\%-45\%) of intra-cloud links by 14\%-33\%.



\end{abstract}

\maketitle


\section{Introduction}

\begin{figure}
    \centering
    \includegraphics[width=\columnwidth]{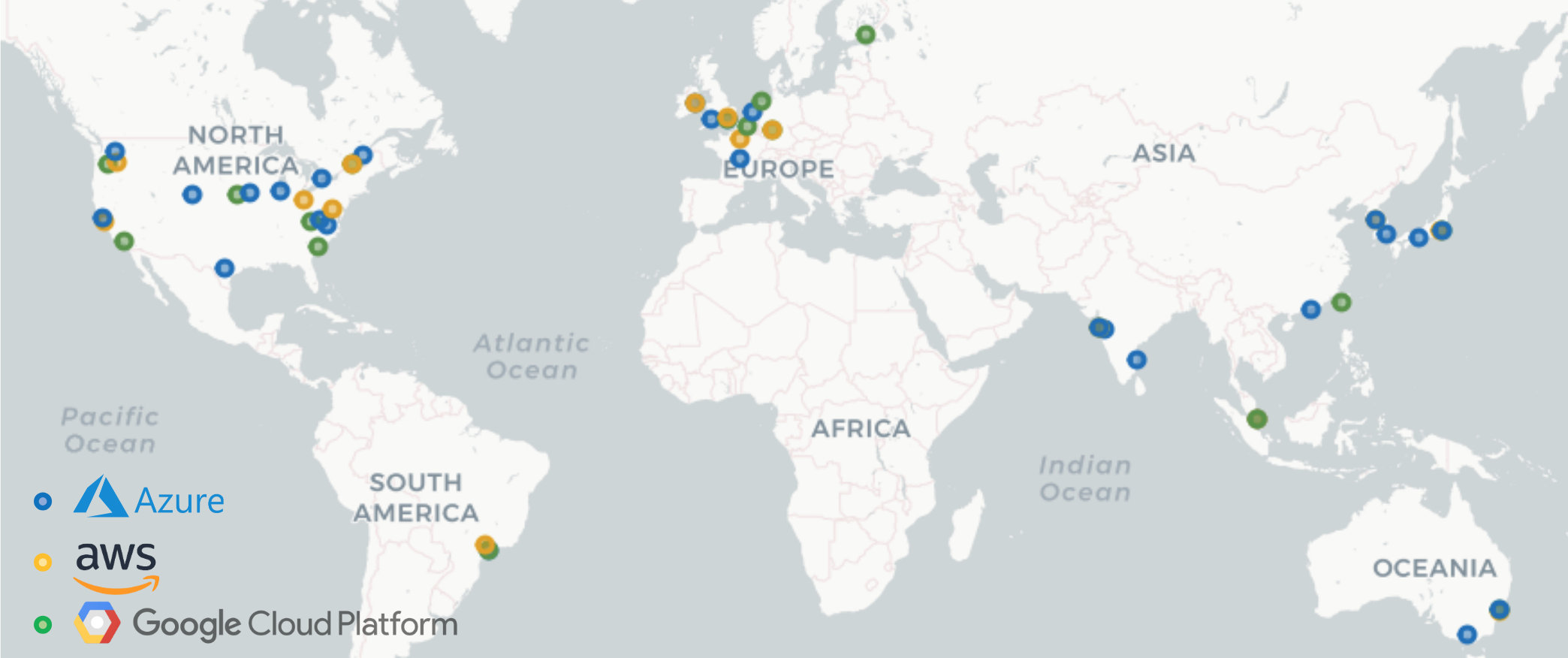}
    \caption{The geographic locations of all the measurement agents used by CloudCast, 59 in total.}
    \label{fig:all_regions}
\end{figure}


Over the past decade, public clouds have emerged as the leading compute and storage platform for hosting major consumer and business applications, as well as primary and backup corporate datacenters. Public clouds are highly popular for many emerging tasks ~\cite{emerging} such 
as databases ~\cite{dbaas}, voice and video analysis, and Machine Learning~\cite{AI1}. Many of these application are distributed across many geographic regions for both resiliency (e.g. secure the continuity of critical data-centers or high-scale interactive application) and the need for proximity to the edge (e.g. self driving cars, IoT, content delivery). Since these tasks are highly data intensive, they need to exchange large quantities of data between the edge and the cloud as well as between public cloud regions.

Public cloud providers\footnote{'public cloud providers', 'cloud providers' and 'clouds' are interchangeably used throughout the paper}, aware of this issue, have spent tens of billions of dollars to improve geographic distribution and data delivery through the buildup of dozens of regional data-centers (termed regions) and high speed long-haul fiber optic networks that connect these regions. For example, in 2018 alone, Google added 5 international regions and 3 intercontinental submarine cables~\cite{google2018}. 
This was is in addition to 11 submarine cables Google has already invested in~\cite{google-fiber}. 
In 2017, Microsoft stated~\cite{azure2017} that it grew its long-haul WAN capacity by 700 percent in 3 years, including the investment in dark fiber and the deployment of a major submarine cable between Virginia, USA, and Bilbao, Spain. 
To demonstrate the size of the global market these public cloud vendors are chasing, Gartner forecasts~\cite{gartner2019} a public cloud services total revenue of over 200B dollars in 2019. The investments in inter-region WAN technologies is not limited to fiber optic infrastructure. Major cloud providers have also reported a development of totally novel SDN based WAN architectures and WAN routing~\cite{b4,b4after,azure2017}.

Just as public cloud computation is a major revenue source for cloud providers, inter-region, and region to Internet data transfer come at a considerable cost to public cloud's customers. Typical charges are in the range of 5-25 cents per GByte of region outgoing transmission, depending on the type of service and region locations 
\cite{aws-price,gcp-price,azure-price}.
However, while public clouds provide computation guarantees (e.g., number and type of CPUs, memory, special hardware, etc.), none of the clouds currently offer any guarantees for data transfer (e.g., in terms of latency, bandwidth or jitter). They also do not provide on-going tracking of network performance and metrics.

Considering the diverse strategies employed by each vendor, and the lack of information in the  way they are operated, it is expected that the performance of each public cloud network and each region may differ significantly. Consequently, there is a need to monitor and compare the performance of the intra-cloud networks along time, so customers can make educated decisions when selecting clouds and regions. A strong multi-cloud trend among public cloud users \cite{multicloud,multicloud1} requires dynamic relocation of workloads across different public clouds. Consequently, it is also important to monitor and compare the networking behavior between different regions of different clouds so that multi-cloud strategies will be optimized for both short and long-terms.

Unlike computation bottlenecks that can be addressed by scaling servers, network latencies within and between clouds always become latencies between, into and out of application servers. 
These in turn impact applications latencies that are crucial for application success.
For example, Amazon reported revenue decreased of 1\% for every 100ms latency~\cite{amazon_latency}.
Google reported that a 100ms slowdown reduces searches by 0.2-0.4\% \cite{brutlag2009speed}.





 
In this paper, we conduct large scale study that attempts to chart the current landscape of cloud data delivery, through the three public cloud leaders -- \textit{AWS} (Amazon web services), \textit{GCP} (Google cloud platform) and Microsoft \textit{Azure}. 
To this end, we constructed \textit{CloudCast} -- a public cloud measurement system, carrying out continuous experimentation throughout all the regions available in the designated clouds, running for almost a year. CloudCast is a world-wide expandable measurement and analysis system, currently collecting data from three major public clouds (\textit{AWS}, \textit{GCP} and \textit{Azure}), 59 unique regions, 1184 intra-cloud and 2238 cross-cloud links (each link represents a direct connection between a specific pair of regions), amounting to 3422 continuously monitored links, and providing active measurements each minute.  
CloudCast is composed of measurement agents automatically installed in each available region, a centralized control, measurement data base, analysis engine and visualization tools. 
\begin{figure*}[t]
    \centering
    \includegraphics[width=1.98\columnwidth]{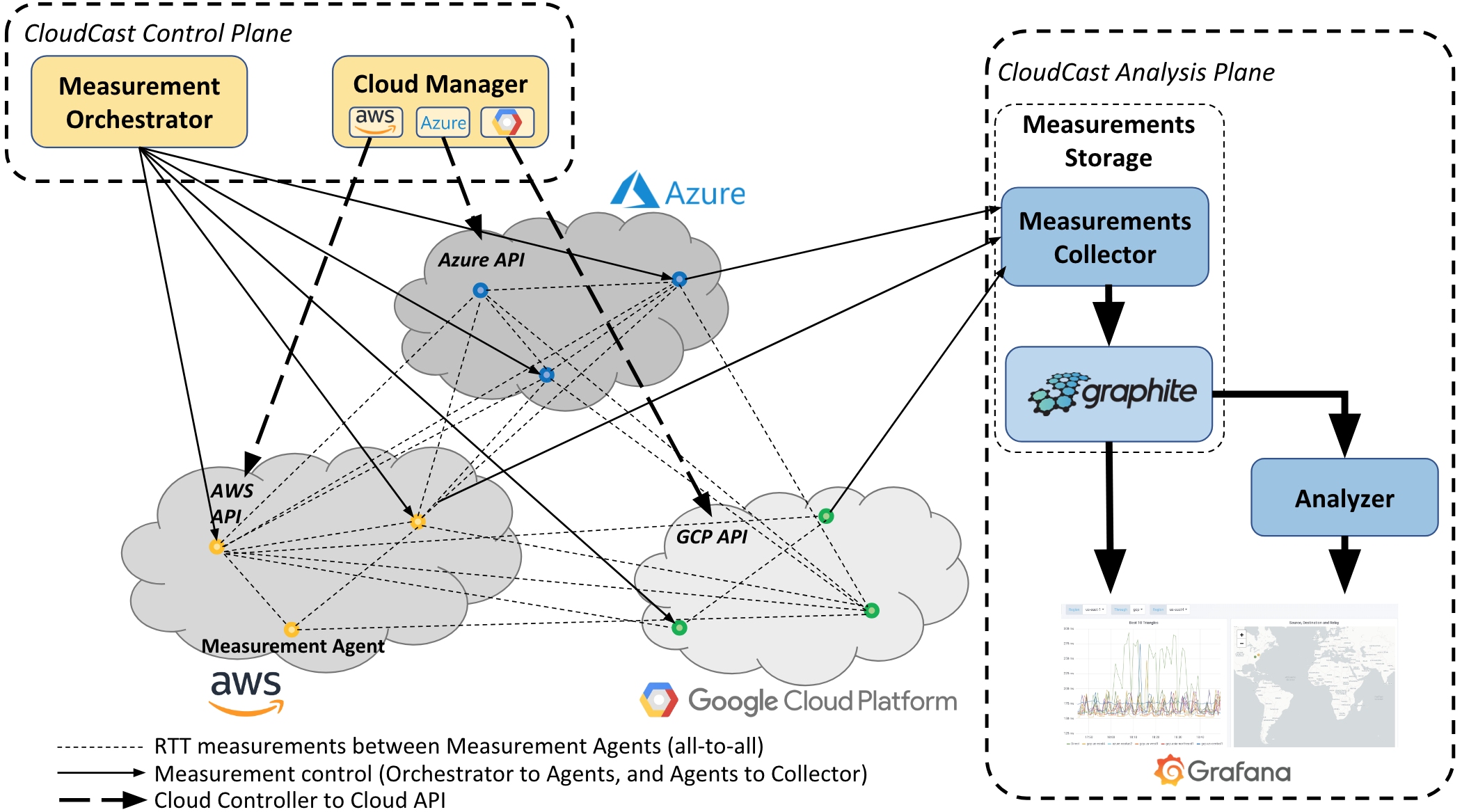}
    \caption{CloudCast components diagram.}
    \label{fig:system_block_diagram}
\end{figure*}



Our main contributions are as follows:
\begin{enumerate}
    \item In \autoref{sec:infrastructure}, we present CloudCast, a comprehensive system for measurement and analysis of public cloud latency. This system includes multiple components, enabling offline examination of the gigs of data collected on a daily basis, as well as real time analysis and visualization, allowing detection of local shortages and other events when they occur.
    \item In \autoref{sec:general_analysis}, we provide a summary of our region to region latency data, and demonstrate the need for more advanced evaluation tools for a fair comparison between clouds.  
    \item In \autoref{sec:similar_links}, we present the \textit{similar links} methodology, which offers a fair comparison between public clouds' networks. We introduce new definitions, and present several interesting results.
    \item In \autoref{sec:triangles}, we introduce a new evaluation tool, termed \textit{\TRIANGLE{}s}. With the help of \TRIANGLE{}s, we are able to compare not only the full roster of our links, but also regions and clouds. Moreover, we present the potential of this tool for solving both short-lived and long-lasting bottlenecks.
\end{enumerate}

\section{Experimental Infrastructure}\label{sec:infrastructure}


We monitored the three leading cloud providers, namely, AWS, GCP and Azure.
The monitoring system is aimed to continuously measure the RTT of all the direct links between all the geographic regions of all the three clouds.
As of October 2018, the total number of links (full mesh) is 3422.
We started to collect and analyze the RTT data almost a year ago, in January 2018.
Our conclusions presented in this paper summarize this one year study of the clouds, and the presented results focus on the most up-to-date measurements performed during October 2018.

\subsection{Cloud Instances}
\label{subsec:cloudinstances}
We utilized a single virtual machine in each geographic region of each cloud.
Some regions allow to create instances in multiple availability zones inside a single region. 
Our tests show that there are no significant link RTT measurement differences between the various zones within the same region. 
Therefore, in order to simplify the analysis and representation and reduce the costs, one of the zones is arbitrarily selected.
In all three clouds we choose a minimal machine type with at least one vCPU and 2~GB of memory, namely, \textit{AWS t2.small}, \textit{Azure A2}, and \textit{GCP n1-standard-1}.
Since the various cloud providers do not offer identical hardware setups, we verified that our test results are not sensitive to the machine level and remain the same if we use a slightly different setup than the minimum mentioned above.
In addition, we have not chosen any privileged network tier, extra service or special hardware.

\subsection{Components}

The process of RTT measurement and analysis is performed using several components, as illustrated in~\autoref{fig:system_block_diagram}. \\

\T{\CLOUDMANAGER.}
The \CLOUDMANAGER~ provides a common interface to initiate cloud instances in various clouds and regions, pick instance flavors, and terminate instances that are no longer needed.
The \CLOUDMANAGER~ is written in Java, and contains specific modules for each cloud, using the APIs and libraries provided by the cloud providers in order to ease such operations when dealing with multiple instances spread across world-wide regions.\\

\T{Measurement Agent.}
The Measurement Agent is a proprietary element developed specifically for our system, and is used to conduct the actual RTT measurements. 
The Measurement Agent wraps multiple simple off-the-shelf measurement tools, adding parallelism, reliability, interface with the other components, and the flexibility to measure the RTT between various clouds, where cloud policies or firewalls may interrupt.
In addition, the Measurement Agent allows to quickly add new regions to the set of regions this agent measures against (termed a mesh), without having to change the setup in every other agent that is already deployed over the different clouds and regions. 
More details regarding the RTT measurements methodology can be found in \autoref{subsec:RTT}.
The agent is written entirely in Python (550 lines), with no dependencies outside the standard distribution of Python~2.7.6.\\

\T{Measurement Orchestrator.}
This is a proprietary module, written in PHP, and running on an Apache2 web server~\cite{apache2}.
The Orchestrator manages the agents: to which IPs to measure the RTT; the measurement rate; the port number to ping (if TCP or UDP is used); and the address of the Measurements Collector to report to.
This flexibility is essential for the ongoing operation and maintenance, as well as quick adaptation to new regions and clouds.


\T{Measurements Storage.} 
This module collects the RTT measurements from the Measurement Agents and stores them for further analysis and visualization.
It is a parent module, made of two sub-modules described below in detail: (1) a storage server employed by \textit{\GRAPHITE{}}~\cite{graphite}, an open source enterprise-ready monitoring tool for storing and retrieving time-series data, and (2) a proprietary \textit{Measurements Collector} that establishes a barrier between the Measurement Agents and the \GRAPHITE{}.\\

\T{Measurements Collector.}
The Measurements Collector is a proprietary server that protects our \GRAPHITE{} server and its stored data.
The isolation of \GRAPHITE{} is needed mainly because it is the bottleneck of our system, and it is a single point of failure.
By design, \GRAPHITE{} does not include protection for data integrity.
For example, \GRAPHITE{} accepts time-stamps reported by clients, allowing them to override historical data, which could leave \GRAPHITE{} instances vulnerable for both deliberate attacks and inadvertent errors.
When old data enters \GRAPHITE{}, it may aggregate the new data with existing data, activating aggregation rules such as min, max, and average, while losing the existing data forever.
Moreover, data in \GRAPHITE{} is organized as key-value pairs, where the key is a unique path. Allowing clients to report directly to \GRAPHITE{}, may expose it to simple key-collision attacks.

Therefore, we protect \GRAPHITE{} by using the Measurements Collector, which is responsible for the following:\\
\TT{Protecting the \GRAPHITE{} server on an IP level} from unauthorized access. This functionality can be easily implemented using the integral firewall of the operating system or the firewall of a hosting cloud. Since the Collector is the only entity allowed to report, it is more practical than allowing the Measurement Agents to go through the firewall.\\
\TT{Throttling the input rate} by verifying that each region reports no more than the desired rate. This is mainly important when running multiple agents per region, for backup or maintenance.\\
\TT{Verifying the location of each reporting Measurement Agent.}
This feature provides region-name consistency and alerts when unknown agent tries to report.\\
\TT{Alerting when a region was not reporting when expected}, catching all kinds of failures quickly.\\
\TT{Generating the measurement time-stamp in a single location,} avoiding any issues with clock synchronization or human errors related to time-zone settings. \\

\T{Graphite.}
We use \GRAPHITE{}~\cite{graphite} version 1.1.4 (September 2018) with a configuration that is very close to the default settings.
The main factors that affect the performance are the number of metrics sent to \GRAPHITE{} per time period, data aggregation and resolution (e.g., calculate and save the minimum per minute), and data retention policy (e.g., keep 1-minute resolution for 12 months). 
We tested several settings in order to reach a setup that copes well with the relatively large number of incoming metrics.
Careful planning is required, since \textit{carbon-cache}, the \GRAPHITE{} component that accepts the metrics, can utilize only a single core CPU, and running multiple carbon-caches may move the bottleneck to disk writes.
Our experience shows that a moderate machine with 1~TB SSD can easily accept 50k metrics per minute, using less than 25\% of a single vCPU dedicated to a single copy of carbon-cache.\\

\T{Grafana.}
Grafana~\cite{grafana} is a leading open platform for monitoring and visualization.
We use it as the system's front end, providing enhanced visualization of time-series data from \GRAPHITE{}.
We provide partial data analysis presented in this paper, as well as up-to-date measurements, with this tool.\\

\T{Analyzer.}
The following describes software developed for an advanced analysis of the measurements, the results of which appear in the following \autoref{sec:general_analysis}, \autoref{sec:similar_links} and \autoref{sec:triangles}.
At first, we considered utilizing the rich collection of built-in functions offered by \GRAPHITE{}. Unfortunately, after we added all the regions, and started to evaluate the quality of links, regions, and clouds, we found that these query-based calculations are too slow for real time analysis.
For example, getting the latest RTT of all the 3422 links takes about 4 seconds, which is reasonable for daily use.
Trying to summarize over the last 5 minutes, or sum up a few links, at times takes more than 30 seconds, causing a timeout (unconfigurable) in \GRAFANA{}.
On the other hand, having our own Analyzer generating the results presented in this paper adds less than a second to the actual read of the data itself.\\


Regarding functionality:\\
\TT{Adding a new region} to the measurement mesh merely requires adding the IP of a newly installed agent to the list maintained by the Measurement Orchestrator. This action triggers the Orchestrator to instruct all other agents to start pinging the new agent.\\
\TT{Measurement rate} is set by default to one RTT measurement per minute among all clouds and regions.\\
\TT{Port number} may vary per target or link. In general, ICMP ping is preferred, as the most inexpensive and common method. When TCP usage is required (e.g., due to firewall rules), the target Measurement Agent listens on a port number known only to the Measurement Orchestrator.\\
\TT{Setting the address of the Measurements Collector} enables to move the Collector during maintenance, replicate the Collector for backup purposes, and run development tests for the Measurement Agent and the Collector.\\

\subsection{RTT Measurement Methodology}
\label{subsec:RTT}
The Measurement Agent wraps reliable known tools: the \emph{ping} utility~\cite{ping,iputils} and \emph{curl}~\cite{curl}. 
The agent uses ping with a single attempt (no retries) and a timeout of one second. 
If a ping fails, as it happens in some clouds (e.g. Azure) and regions due to firewall blocking, then the agent makes an attempt to perform TCP connect instead and measures the connect time of curl.
We verified that curl's reported connect time is identical to ping, when both are successful.
In addition, to bypass possible blocks of known TCP ports, and allow agents to run as non-root, the agent listens on a high port number (above 5000), reports it to the Orchestrator, which in turn configures the TCP connect attempts from other agents.

Each agent completes a single round of pings to all regions in all clouds within a very short interval of less than 5 seconds.
Such an operation requires careful parallel run of pings, taking into account the potential switch to curl instead of ping.

\section{Basic Analysis}\label{sec:general_analysis}

What makes a cloud's network "good"? Naturally, an intra-cloud link that provides relatively low and stable latency all day long, is a good link. 
Unfortunately, as we show in this paper, such links rarely exist. Moreover, each cloud's network is made of many links, each possibly demonstrating different characteristics, thus making our cloud profiling assignment highly challenging. 

In this section we evaluate the dynamic performance of the three clouds' networks (AWS, GCP and Azure) through the variance of their RTT link measurements. 
We separate this study to the study of intra-cloud and inter-cloud connectivity. 
The intra-cloud part helps us in evaluating the performance of the proprietary network established by each public cloud provider to connect its own regions. 
The inter-cloud portions enables us to examine the level of connectivity across different cloud providers.

\subsection{RTT is Very Dynamic Across Different Time Scales}\label{subsec:dynamic}

Monitoring clouds' networks for almost a year, we have observed over most links significant fluctuations in RTT, in several patterns as presented in \autoref{fig:patterns}: 

\begin{figure*}[!t]
    \centering
    \begin{subfigure}{.49\textwidth}
  \centering
 \includegraphics[width=0.90\textwidth]{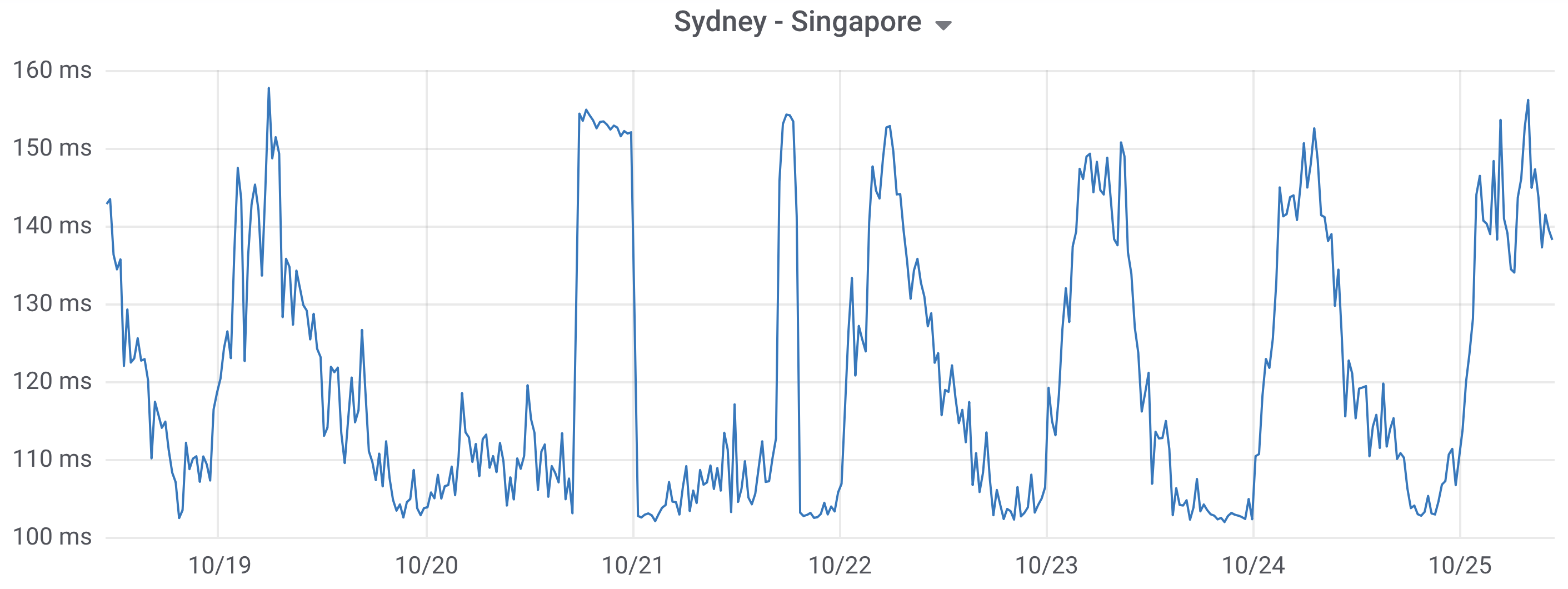}
    \caption{Daily repetitions.}
    \label{fig:pattern_daily_changes}
\end{subfigure}
\begin{subfigure}{.49\textwidth}
  \centering
 \includegraphics[width=0.90\textwidth, trim={0cm 0cm 0cm 0cm}, clip]{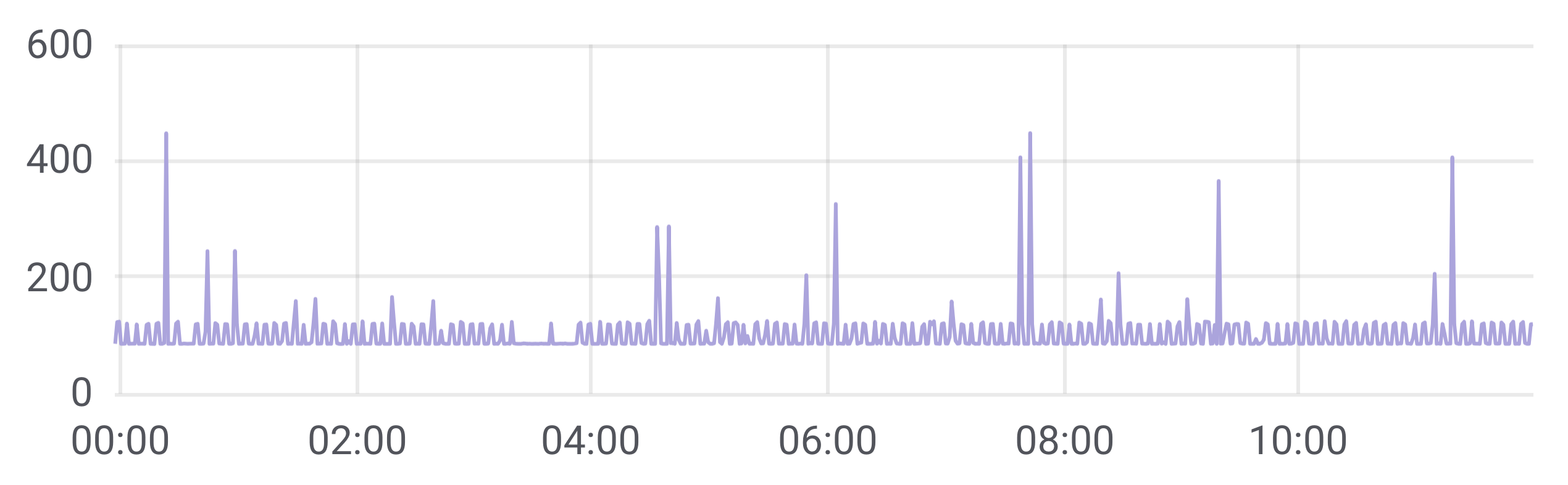}
    \caption{Short-lived peaks.}
    \label{fig:pattern_peaks}
\end{subfigure}
\begin{subfigure}{.49\textwidth}
  \centering
 \includegraphics[width=0.90\textwidth, trim={0cm 0cm 0cm 0cm}, clip]{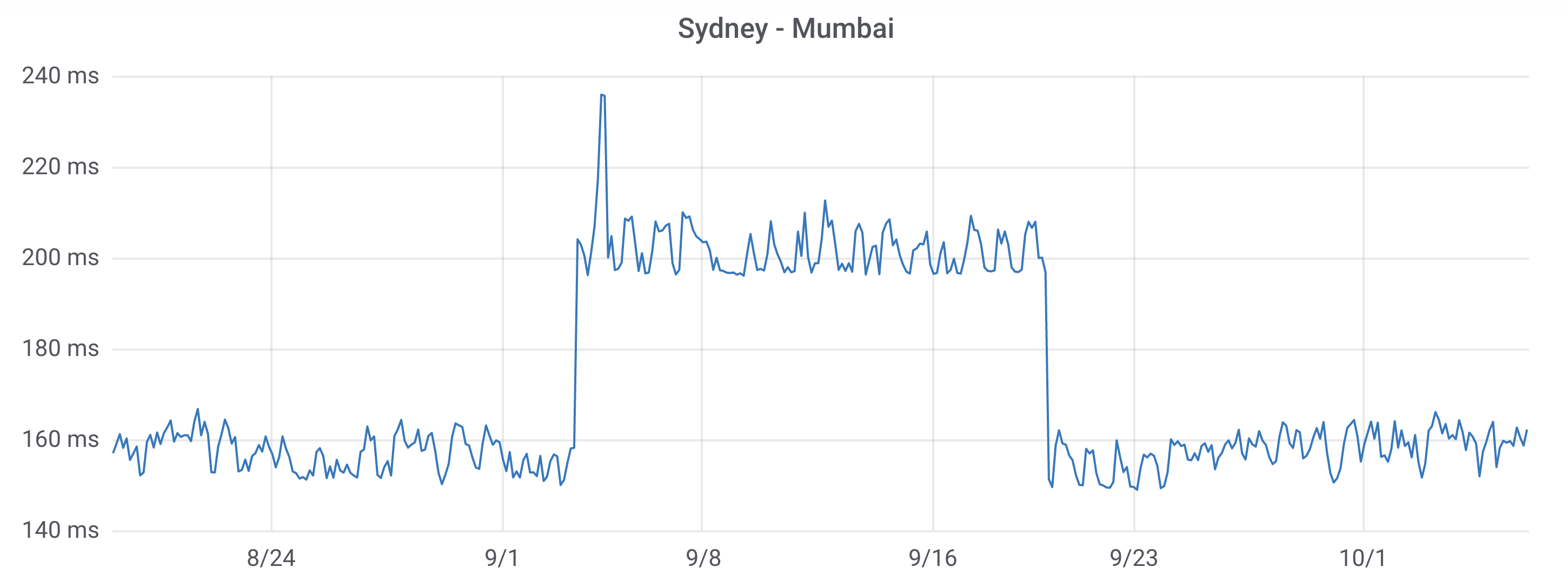}
    \caption{Shortages that last for hours.}
    \label{fig:pattern_shortage}
\end{subfigure}
\begin{subfigure}{.49\textwidth}
  \centering
 \includegraphics[width=0.90\textwidth, trim={0cm 0cm 0cm 0cm}, clip]{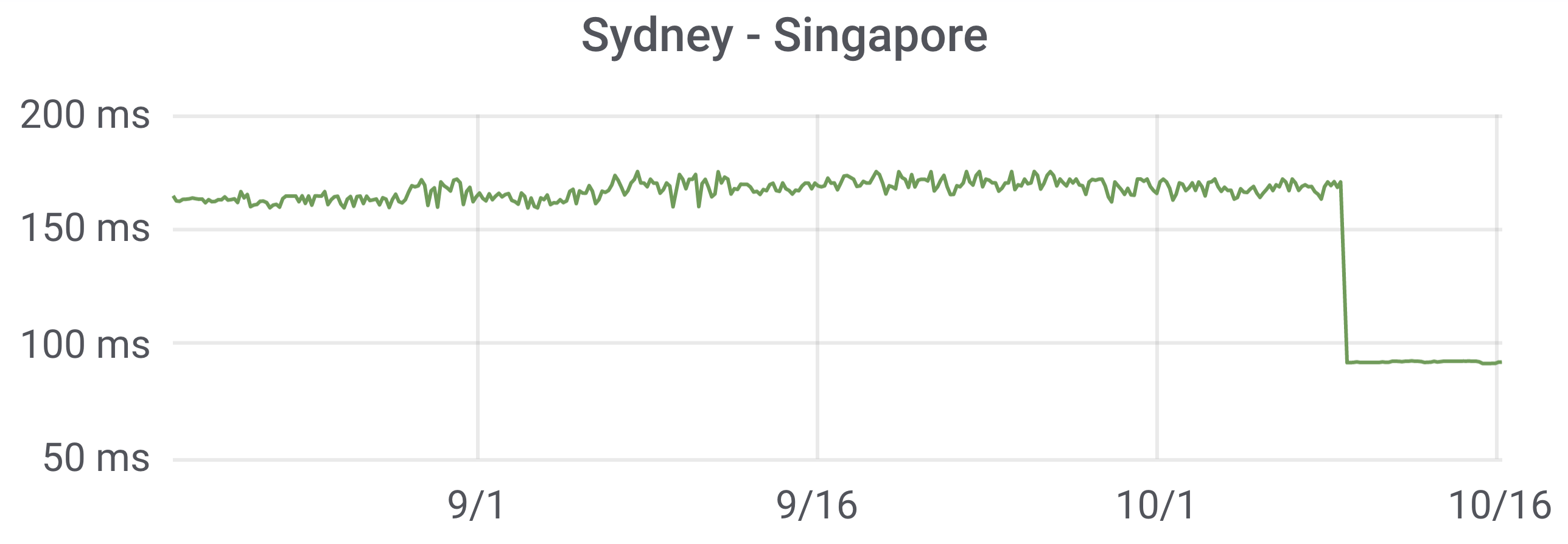}
    \caption{Routing or physical layer enhancements.}
    \label{fig:pattern_routing_enhacements}
\end{subfigure}
    \caption{Examples of patterns of RTT changes in various clouds, regions and time periods.}
    \label{fig:patterns}
\end{figure*}

\T{Daily repetitions.} RTT oscillates in a 24-hours pattern, which we believe can be contributed to network load changes along the day. For example, a sine-wave like pattern presented in \autoref{fig:pattern_daily_changes}.\\ 
\T{Peaks.} Significant \textit{short-lived} increases or \textit{spikes} in RTT, that repeat several times a day in an inconsistent pattern.
A short-lived peaks pattern is illustrated in \autoref{fig:pattern_peaks}.\\
\T{Shortages.} Problems in specific links that suddenly switch to a significantly higher RTT. 
We found that a shortage usually affects one or two regions, along with most of the links connecting them to others, and usually lasts a few hours to a few days, before going back to the shorter RTT.
It may be related to maintenance, power outages, or temporary outages in the fiber optic layer (\autoref{fig:pattern_shortage}). \\
\T{Routing or physical layer enhancements.} Improved RTT results, probably due to enhancements in routing or physical layer in and between regions, as shown in \autoref{fig:pattern_routing_enhacements}.
A routing enhancement affects very specific links, and lasts for at least few months.\\


\begin{figure*}[t]
    \centering
    \includegraphics[width=\textwidth]{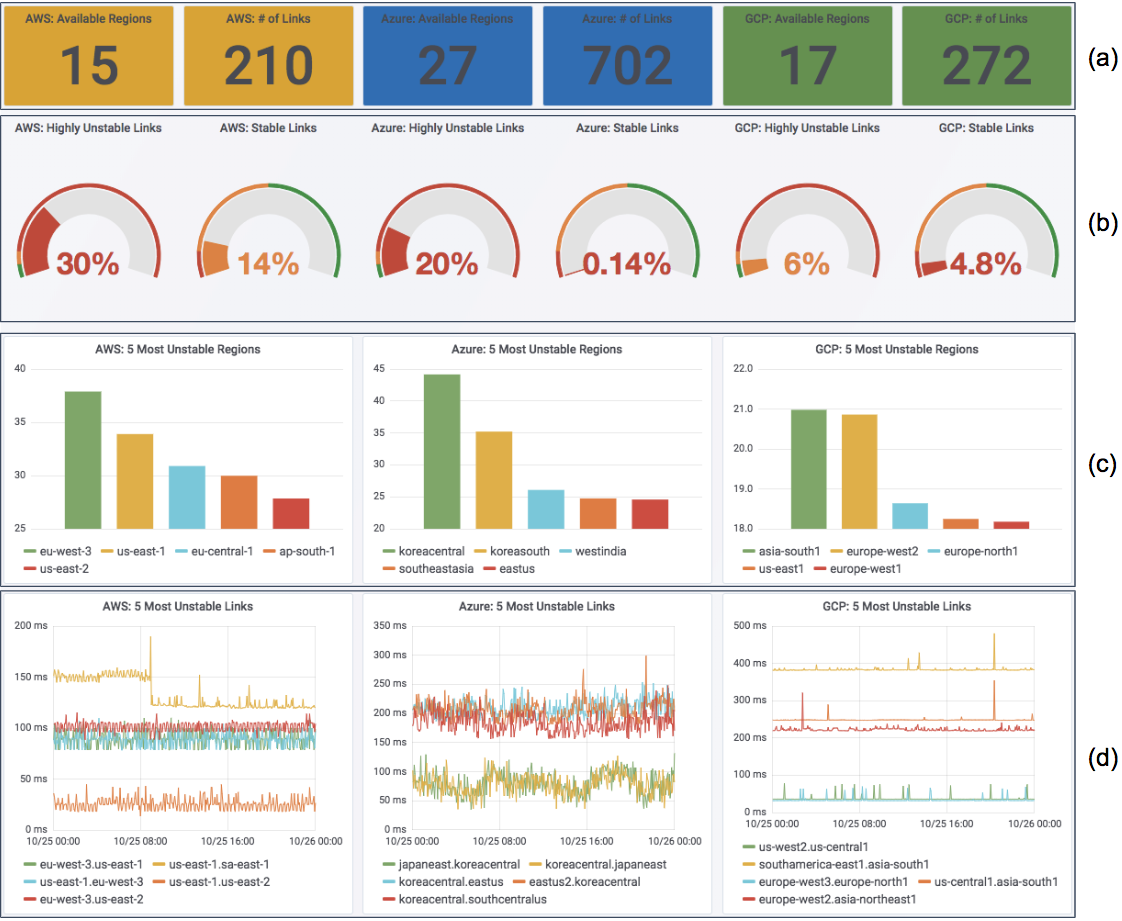}
    \caption{Public clouds vary in even the most basic parameters. This is a portion a Graphana dashboard dedicated to intra-cloud links. The data displayed was collected over a 24 hour period during October 25 2018. Each of the three columns showcases data stemming from a particular cloud - AWS, Azure or GCP.
    The row marked (a) shows the number of available regions and intra-cloud links in each public cloud.
    The gauges in section (b) indicates the percentage of stable and highly unstable links. The most unstable regions are shown in (c).
    The graphs of the most unstable intra-cloud links are plotted in (d).}
    \label{fig:intra_dashboard}
\end{figure*}

\begin{figure*}
    \centering
    \includegraphics[width=\textwidth]{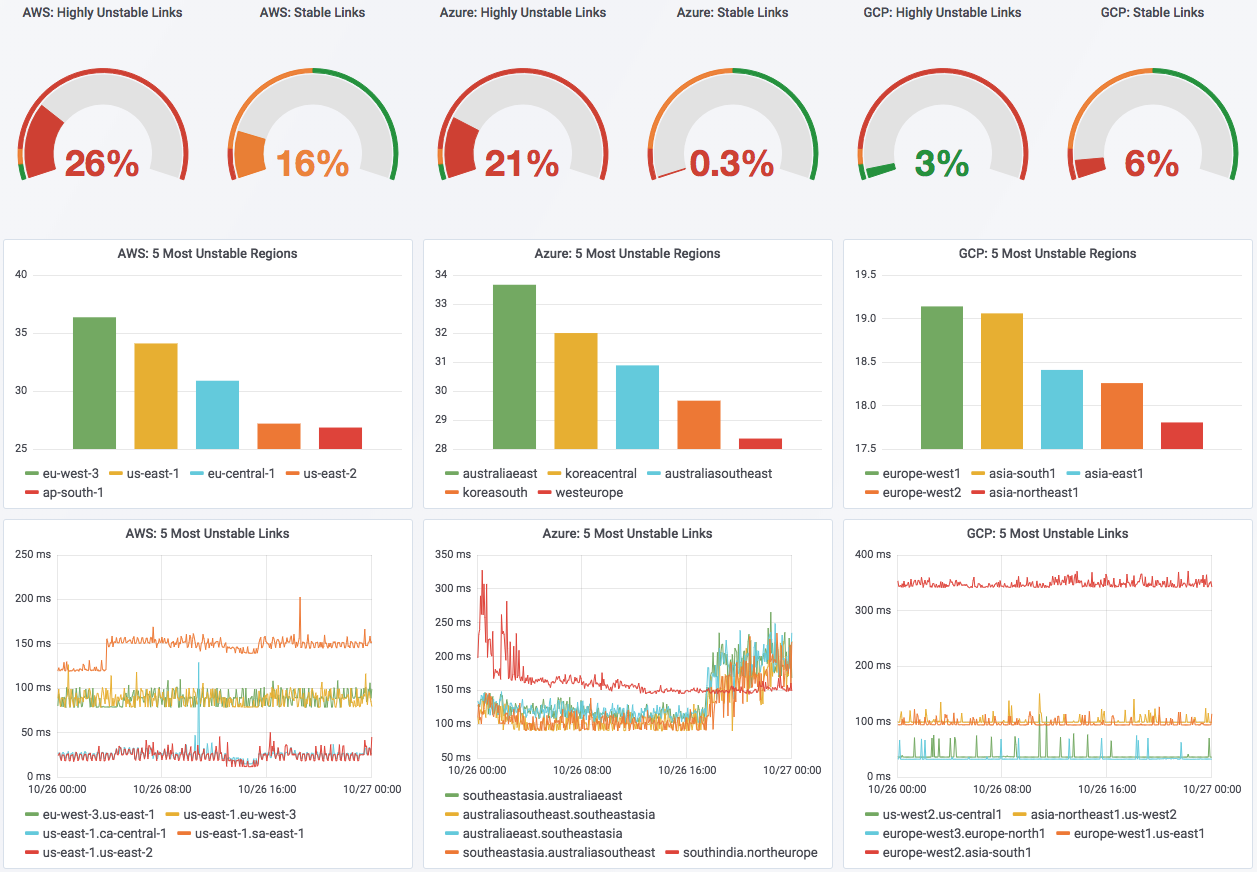}
    \caption{Public clouds connectivity behavior changes through time. This is a portion of the dashboard presented in \autoref{fig:intra_dashboard}, populated with the data from the following day - October 26, 2018.
    Even when examining consecutive days, changes can be easily detected.}
    \label{fig:intra_dashboard_2}
\end{figure*}

\subsection{Public Clouds are Not Created Equal}


When we began our survey with the most basic parameters, namely RTT values and standard deviation comparison, we were surprised to find an ongoing significant diversity in results. 
\autoref{fig:intra_dashboard} showcases this variability in intra-cloud links. Throughout the paper, we demonstrate this point in several ways. 

This comparison can be unfair, penalizing public cloud providers for deploying datacenters in less 'popular' regions that may not enjoy upgraded infrastructure. To avoid this bias, we have also conducted a specialized analysis, focusing on the more popular geographic locations, presented in \autoref{sec:similar_links}.

\subsection{Intra-Cloud Links}\label{subsec:intra}
Analyzing intra-cloud links allows us to pinpoint behaviors specific to each single cloud. 
In some ways, the data showcased in~\autoref{fig:intra_dashboard} is very representative of what we have observed for several months. We define a \textit{stable link} to be a link with standard deviation of less than one millisecond, and a \textit{highly unstable link} to be one that its standard deviation is bigger than ten milliseconds. 
Similarly, we look at the standard deviation of all links going in or out of a region, and consider that number to be an estimator of a region's quality, where high numbers mean (relatively) \textit{unstable region}.

\T{Most links exhibit large standard deviation.} 
The number of stable links was only 43 out of the 1184 intra-cloud links, while there were 220 highly unstable links.

\T{The number of highly unstable links varies from cloud to cloud.} 
During last year we have observed that GCP usually has the smallest number of unstable links among the three clouds. 
In this particular sample day of~\autoref{fig:intra_dashboard}, AWS had more unstable links than Azure, however, from our experience, this trend fluctuates on a daily basis.

\T{Number and location of unstable regions differs between clouds.}
One or two regions in Azure usually display a very high standard deviation (40-50ms), while the rest have high to moderate standard deviation. 
AWS, for the most part, has a few regions with very high standard deviation. 
Regions in GCP, for the most part, display only high to moderate standard deviation, with no region having a very high standard deviation. 
The location of unstable regions is ever changing, and for the most part differs from cloud to cloud.

To better illustrate how cloud behavior is developing over time, we present two dashboards:
\autoref{fig:intra_dashboard} and \autoref{fig:intra_dashboard_2}, showing data for two consecutive days. 
While each cloud has a typical overall behavior over time (as described above), changes are quite noticeable even when examining short time periods.
For example, in Azure, the list of the top five most unstable links (in terms of standard deviation) has significantly changed. 
In AWS, we see how East US to South America East link (us-east-1.sa-east-1) experiences a sudden  decrease in RTT the first day, while experiencing an increase the second day.

\subsection{Cross-Cloud Links}
In this subsection, we bring examples of our finding using the measurements collected on October 25th, 2018. These are inline with the trends we have seen while analyzing months of data.

\T{The percentage of links with high standard deviation is only moderately higher than in intra-cloud.} In our sample data, a mere 4\% of all the cross cloud links were stable, while 23\% were highly unstable (see \S\ref{subsec:intra}). As these links are expected to traverse more than one cloud, and also to cross the Internet, we would expect to see a much more "noisy" behavior. To our surprise, these numbers are on par with the statistics of the intra-cloud links, with highly unstable links constituting 18\% of all links, and stable links making up  close to 4\%.


\T{Connectivity between AWS and GCP is the most stable.} On the day our sample was collected, 10\% of the links between AWS and GCP were stable. 18\% were highly unstable.

\T{Connectivity between AWS - Azure and GCP - Azure is more unstable.} Between AWS and Azure, 35\% of the links were highly unstable, and only 4\% were stable. Between GCP and Azure, while only 14\% were highly unstable, the percentage of stable links was only 1.1\%.





\section{Similar Links}\label{sec:similar_links}
\subsection{Definitions}
AWS, GCP and Azure deploy their own fiber-optics and routing infrastructure to connect their datacenters, which are deployed over different geographic regions. 
In order to better evaluate and compare the performance of the proprietary underlying network technologies between different cloud providers (in terms of RTT), we perform a 'common-denominator' comparison between the RTT of the intra-cloud links between all the cloud providers over their mutual geographic regions -- termed \textit{similar links}. 
This methodology constitutes a fair comparison, as using this methodology we never penalize cloud vendors that deployed more datacenters in unique and possibly under-served geographic regions that may lack fiber optic infrastructure compared to more 'popular' geographies.
\\
\begin{definition}
Let $A,B$ be datacenters in clouds $C_1,C_2$ respectively. We say that $A,B$ are \textit{adjacent} iff:
\begin{enumerate}
    \item $C_1 \neq C_2$
    \item  $A$ and $B$ are located in the same geographic region. 
\end{enumerate}
\end{definition}

A list of adjacent datacenters can be found in \autoref{tab:similar_locations}. In total, we identified 11 unique hubs of adjacent datacenters shared by all three clouds reviewed in this paper. 

\begin{definition}
Let $R^{(C,a,b)}$ be a route in cloud $C$ such that datacenter $a$ is the source of the route, and datacenter $b$ is its
destination. We say that $R^{(C_1,a,b)}$ and $R^{(C_2,d,e)}$ are \textit{similar links} iff:
\begin{enumerate}
    \item $C_1 \neq C_2$
    \item $a$ and $d$ are adjacent
    \item $b$ and $e$ are adjacent
\end{enumerate}
Note that route similarity is a transitive relation.
\end{definition}

\begin{table}[t]
\begin{center}
\resizebox{\columnwidth}{!}{
\begin{tabular}{ l c c c }

\textbf{Geographic Region} & \textbf{AWS} & \textbf{Azure} & \textbf{GCP} \\
\hline \hline
US East &   us-east-1   &   eastus  & us-east4\\
US West &   us-west-1   &   westus  & us-west2\\
US North West & us-west-2   &   westus2 &   us-west1 \\   
Canada East  &   ca-central-1    &   canadaeast  &   northamerica-northeast1 \\
South America East  &   sa-east-1   &   brazilsouth &   southamerica-east1 \\
UK South    &   eu-west-2   &   uksouth &   europe-west2 \\
Europe Central &   eu-central-1    &   westeurope  &   europe-west4 \\
India West  &   ap-south-1  &   westindia   &   asia-south1 \\
Singapore   &   ap-southeast-1  &   southeastasia   &   asia-southeast1 \\
Japan East  &   ap-northeast-1  &   japaneast   &   asia-northeast1 \\
Australia South East    &   ap-southeast-2    &   australiaeast   &   australia-southeast1 \\
\end{tabular}
}
\caption{List of clouds' data-centers with similar geographic regions}
\label{tab:similar_locations}
\end{center}
\end{table}

We compare RTT between similar links. 
Altogether, this category contains 110 links. 
This category illustrates the variability in RTT and behavior among the three major public cloud providers. 

\subsection{Best and Worst Links Statistics}

\begin{figure*}
    \centering
    \begin{subfigure}{.48\textwidth}
        \centering
        \includegraphics[width=0.9\columnwidth]{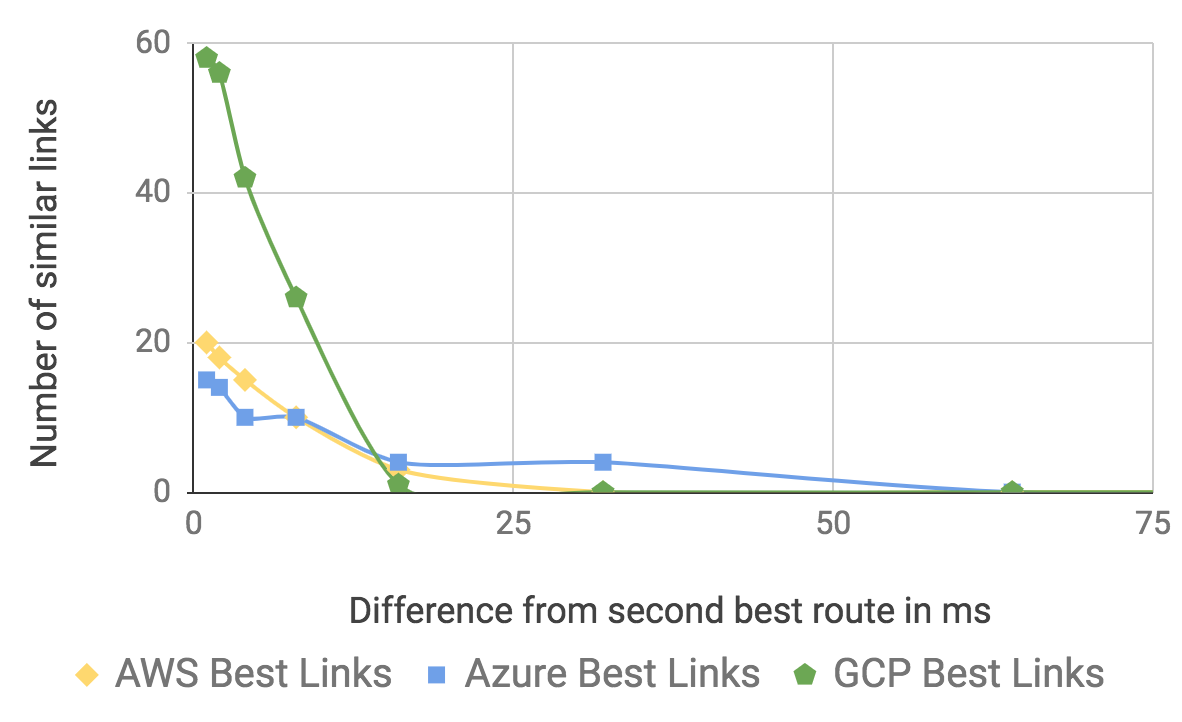}
        \label{fig:best_similar_links}
        \caption{Best similar links}
    \end{subfigure}
    \begin{subfigure}{.48\textwidth}
        \centering
        \includegraphics[width=0.9\columnwidth]{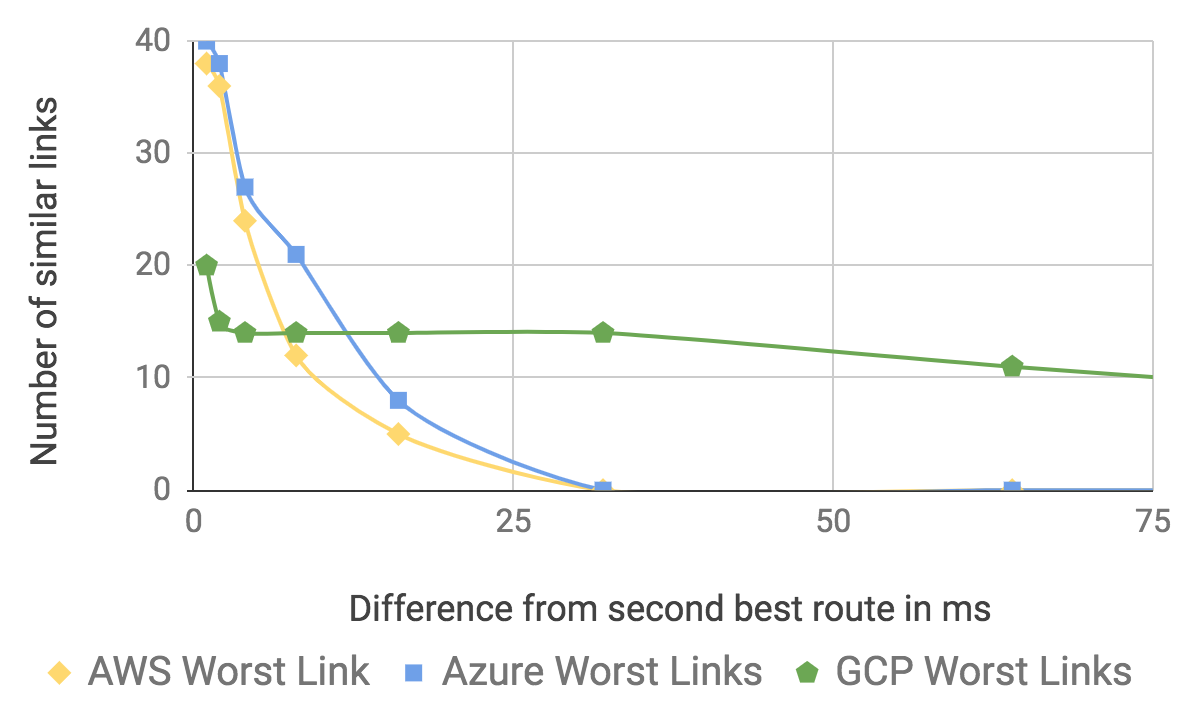}
        \label{fig:worst_similar_links}
        \caption{Worst similar links}
    \end{subfigure}
    \caption{Comparing the best and worst performing similar links in each cloud cloud. The Y axis is the number of links. The X axis is the difference in RTT (ms) from the second-best route. 
    Clouds vary in number links per category, as well as in the RTT improvement or decrease from second-best route.
    }
    \label{fig:similar_routes_cdf}
\end{figure*}

\autoref{fig:similar_routes_cdf} summarizes the number of similar links each provider excelled at, and those that had the worst results throughout the month of September 2018. 
The X axis shows by how much the RTT of that link differs from the second best link of that similar route.

AWS has a low count of best links and a high count of worst links. 
In terms of the least-performing links, the different between those and the second-best links drops quickly, the largest difference being 16ms, and applies only to 5 links.
Azure had the least best-performing similar links, however, the RTT along these links is much better than the second-best route than the other two cloud providers.
GCP has the largest number of best links, though most of those are better than the second-best route by only 4ms. In addition, while GCP has the lowest number of worse-performing links, the different between these and their second-best counterparts is much higher, some are higher than 75ms.

\subsection{Correlation Between RTT and Standard Deviation}

\begin{figure}[]
    \centering
    \includegraphics[width=\columnwidth]{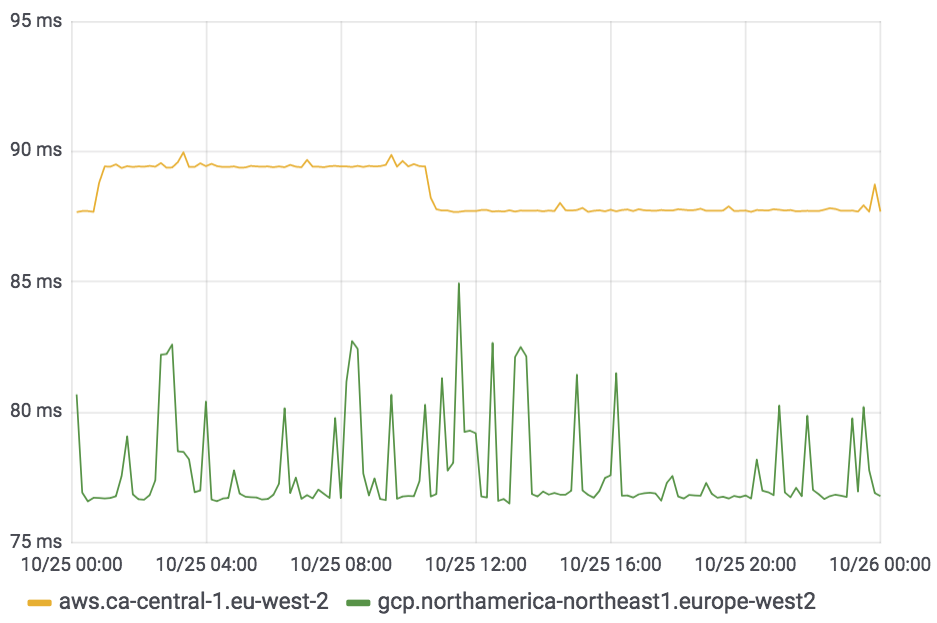}
    \caption{Lower standard deviation does not imply lower RTT. Presenting AWS and GCP links between Canada East and UK South. The data was collected during October 25th 2018. While the average RTT of the GCP link is lower than the AWS link by 10ms, the standard deviation of the GCP link is much larger.}
    \label{fig:similar_links_1}
\end{figure}

To some consumers and applications, such as VoIP, a stable RTT is preferable over a lower-but-noisy RTT, while other may rather utilizing links that provide the lowest average RTT.
We set out to examine whether public clouds are able to provide links that posses both properties.
To do so, we compared results collected from similar links. 
As shown in~\autoref{fig:similar_links_1}, having a low standard deviation does not imply a lower RTT. 

It is true that some links, as mentioned in~\autoref{sec:general_analysis}, exhibit a large standard deviation on a regular basis. This observation may cause customers to become apprehensive, however, this trait is not indicative of a lower average RTT. In the case of Azure, for example, we constantly see links that perform much better than others, and as can be seen in~\autoref{fig:similar_routes_cdf}, at times by larger numbers, even though they may seem to have a larger standard deviation.

\subsection{Beyond Similar Links}
Examining similar links, as we have done in this section, has provided important insights into the ability to provide a fair cloud comparison.
This methodology, however, is limited by definition to 110 links out of 3422 possible options. 
In the following section, we will introduce a new tool, that does not only allow us to include more links in our comparison while maintaining fairness, but also offers a possibility of improving latency through a simple technique.


\section{\TRIANGLE{}s} \label{sec:triangles}

In this section we present a novel tool, termed \textit{\TRIANGLE{}}, which offers a fair characterization of public clouds connectivity.
The \TRIANGLE{} tool provides the ability to quickly present the current state of each link, region, and cloud, while taking into account the unique geographic deployment of each cloud.
Furthermore, this tool introduces a possible solution to temporary and permanent bottlenecks, and helps to estimate when it might be beneficial for a client to route traffic through more than one cloud.

\subsection{Cloud Instances as \RELAYUC{}s}\label{subsec:triangles}




In Euclidean geometry, the triangle inequality states that for any triangle, the sum of the lengths of any two sides must be equal or greater than the length of its' remaining side. In networking, however, this inequality does not always hold. In other words, when triangle inequality does not hold, routing through an additional relay (cloud's region) may lead to lower latency.  
The results we obtained in \autoref{sec:general_analysis} made us wonder, how many times this inequality holds in the case of public clouds? This is an interesting question, as we assume that the underlying infrastructure of each cloud provider attempts to route traffic in the best possible way over its own network. However, operating across cloud providers and sometimes over Internet peering points have the potential of achieving better results through an indirect route. To answer this question, we began monitoring a phenomenon which we define as a \textit{\TRIANGLE{}}. 
\begin{definition}
Let $\mathcal{R}$ be the group of all available regions. For $R_s, R_d \in \mathcal{R}, s\neq d$, let $D_T(R_s, R_d)$ be the minimal RTT measured from $R_s$ to $R_d$ in a time interval T. We say that $(R_s, R_r, R_d)$ is a \textit{\TRIANGLE{}} iff $s\neq d \neq r, R_r \in \mathcal{R}$ and:
$$D_T(R_s, R_d) > D_T(R_s, R_r) + D_T(R_r, R_d) $$
We refer to $R_s$ as the \textit{source}, $R_d$ as the \textit{destination} and $R_r$ as the \textit{\RELAY{}}.
\end{definition}

Intuitively, \TRIANGLE{}s are instances in which the triangle inequality does not hold.

We now turn to define the term \textit{\BESTTRIANGLE{}} and \textit{\BESTRELAY{}}:
\begin{definition}
Let $(R_s, R_r, R_d)$ be a \TRIANGLE{}. We say that $(R_s, R_r, R_d)$ is a \textit{\BESTTRIANGLE{}} and $R_r$ is a \textit{\BESTRELAY{}} iff $s\neq d \neq r \neq i$ and:
$$\forall R_i \in \mathcal{R}, D_T(R_s, R_i) + D_T(R_i, R_d) \geq D_T(R_s, R_r) + D_T(R_r, R_d) $$
\end{definition}

\subsection{\TRIANGLEUC{}s in Practice}

The notion of \TRIANGLE{} is in fact, very practical: if a triangle is found to improve the delay significantly, meaning $D(R_s, R_d) >> D(R_s, R_r) + D(R_r, R_d) $, one may route the traffic from $R_s$ destined to $R_d$, through a \RELAY{} in $R_r$, in order to experience the smaller RTT this indirect route offers. 
Although such 2-hop routing may be subjected to higher public cloud bandwidth charges, it is practical, and will add only a negligible latency in the \RELAY{}, as shown in several previous works~\cite{pucha2005overlay,snellman2015mobile}.

\begin{figure*}[t]
    \centering
    \begin{subfigure}{.33\textwidth}
        \centering
        \captionsetup{width=.9\linewidth}
        \includegraphics[width=0.95\textwidth, trim={0cm 0cm 0cm 0cm}, clip]{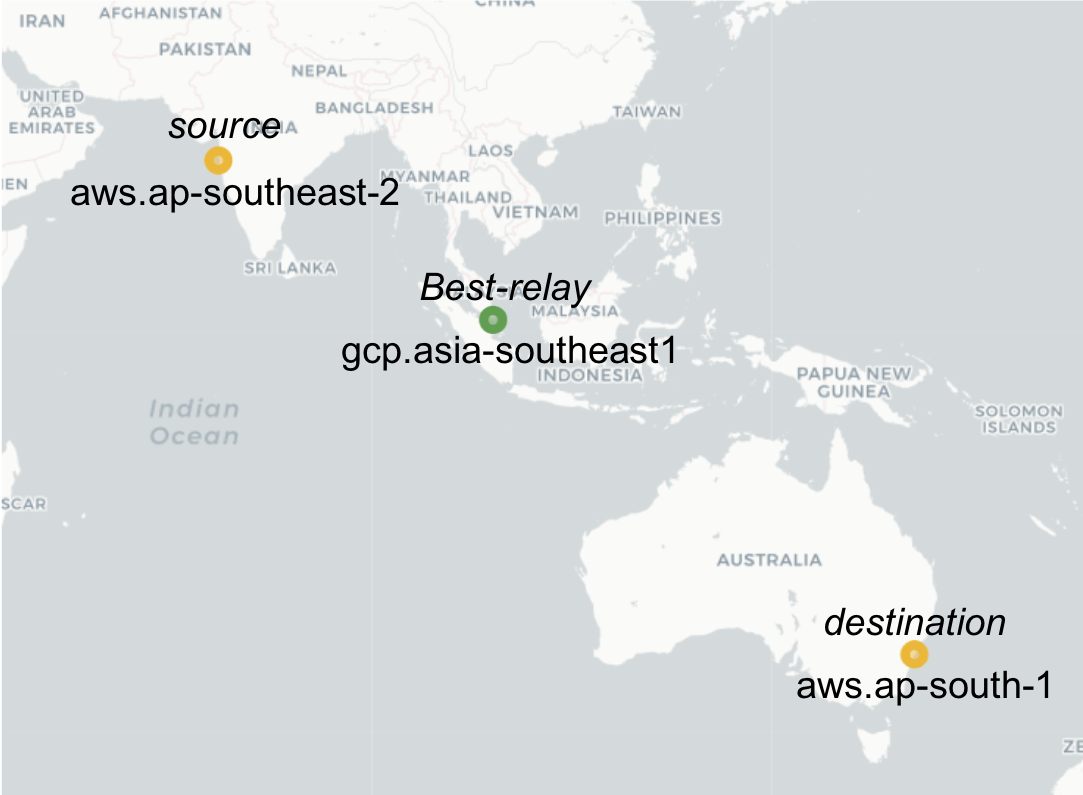}
        \caption{\BESTTRIANGLEUC{} map, where the \BESTRELAY{} is GCP in Singapore, and it is visually located at the center of the picture, between the source and the destination regions.}
        \label{fig:triangle_map}
    \end{subfigure}
    \begin{subfigure}{.33\textwidth}
        \centering
        \captionsetup{width=.9\linewidth}
        \includegraphics[width=0.95\textwidth, trim={0cm 0cm 0cm 0cm}, clip]{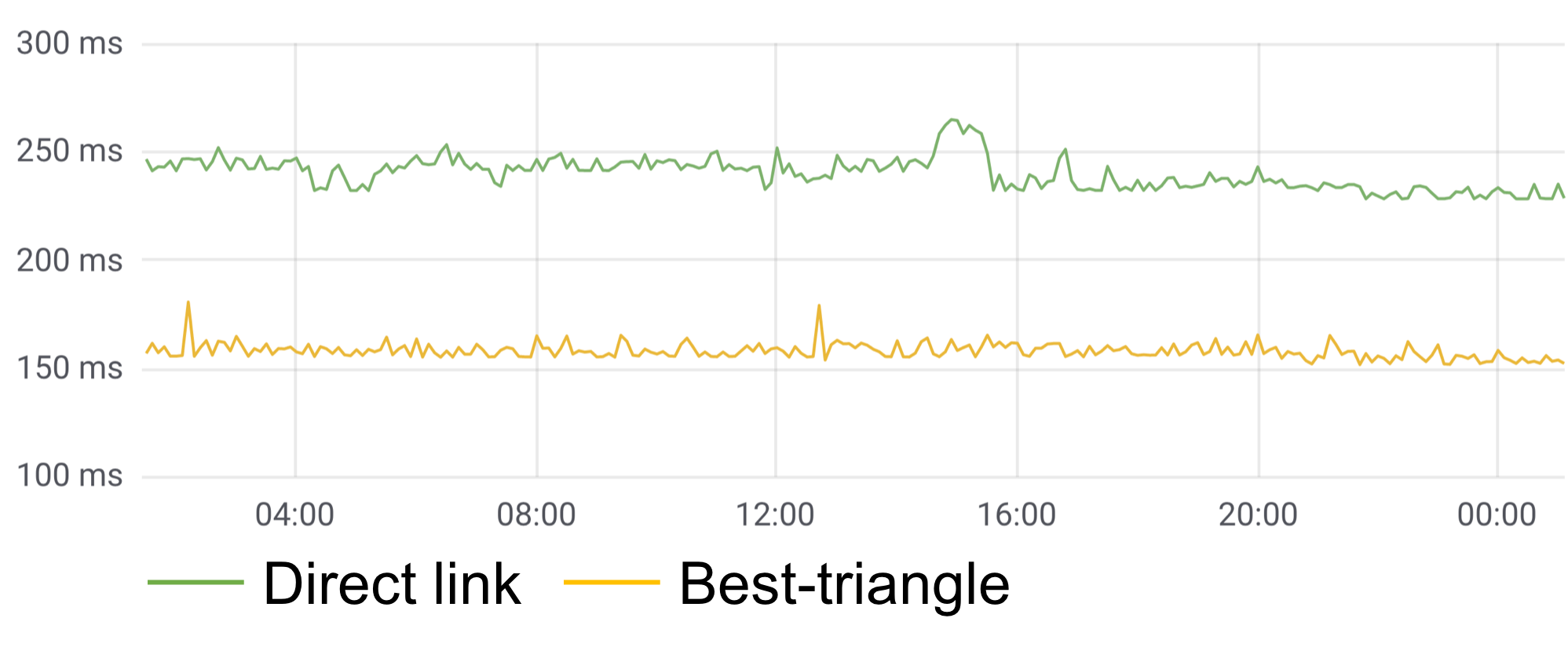}
        \caption{\BESTTRIANGLEUC{} 24-hours graph, showing the RTT of the direct link (upper line, average 239.83~ms) vs. the RTT of the \BESTTRIANGLE{} (lower line, average 158.07~ms).}
        \label{fig:triangle_graph}
    \end{subfigure}
    \begin{subfigure}{.33\textwidth}
        \centering
        \includegraphics[width=0.90\textwidth, trim={0cm 0cm 0cm 0cm}, clip]{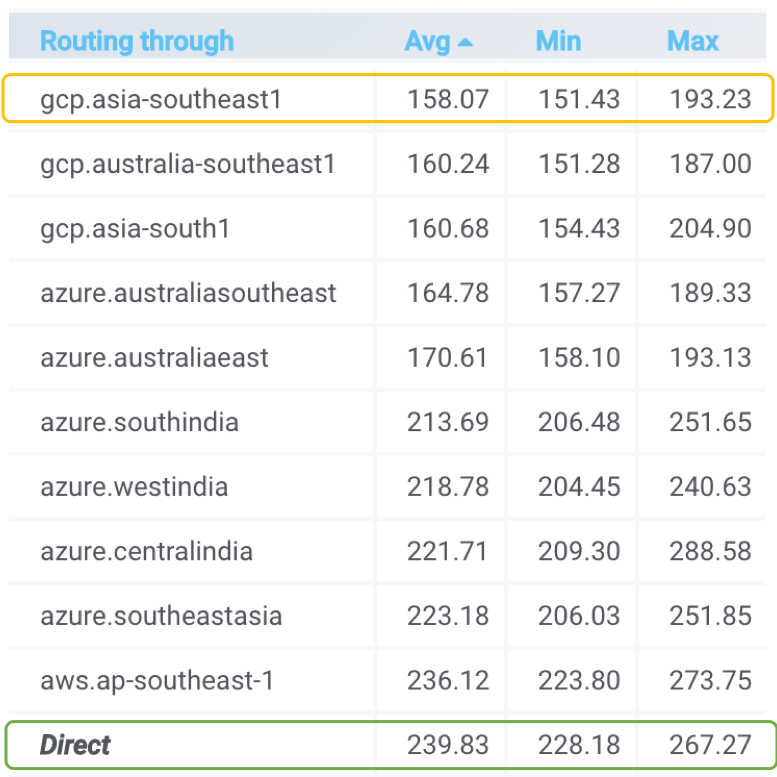}
        \caption{List of top potential \TRIANGLE{}s (\RELAY{}s) for this source and destination in the 24-hours snapshot, sorted by average overall latency. The highlighted \RELAY{} at the top is the \BESTRELAY{} of the \BESTTRIANGLE{} shown in the map and graph.}
        \label{fig:triangle_table}
    \end{subfigure}
    \caption{An example of \TRIANGLE{}s and \BESTTRIANGLE{}. The source and destination instances are both AWS, located in Mumbai and Sydney, accordingly. The \BESTTRIANGLE{} consists of a \BESTRELAY{} of a GCP instance located in Singapore.}
    \label{fig:triangle_example}
\end{figure*}

\autoref{fig:triangle_example} demonstrates the concept of \TRIANGLE{} and \BESTTRIANGLE{} via an example of a specific link. 
In this example, the source and the destination are AWS's regions in Mumbai India and Sydney Australia regions.
Our measurements and calculations found that the \BESTTRIANGLE{} contains a \RELAY{} in GCP's Singapore region (i.e., \BESTRELAY{}), as illustrated in \autoref{fig:triangle_map}.
\autoref{fig:triangle_graph} illustrates the RTT measured in the direct link from source to destination (the upper line) vs. the RTT of the \BESTTRIANGLE{} that is the sum of the two links going from destination to \RELAY{} and from the \RELAY{} to the destination.
\autoref{fig:triangle_table} shows that in this case we found 9 more \TRIANGLE{}s that are significantly better than the direct link.
These triangles are not the \BESTTRIANGLE{}, but they also result in a significantly lower RTTs than the direct link, by adding a single \RELAY{} in between.

\subsection{Unveiling the Potential of \TRIANGLEUC{}s}

In our live system, we calculate triangles in real-time every minute, for a time interval $T$ of the recent 5 minutes. 
To this end, we take a complete snapshot of RTT measured in all 3422 links during the recent 5 minutes, take the minimum RTT of the 5 RTTs (one per minute) in each link, and then check each of the 57 potential \RELAY{}s to see if it establishes a \TRIANGLE{}.
Altogether, we calculate 195054 optional \TRIANGLE{}s every minute, ending up with about 2000 links having at least one \TRIANGLE{}.

The high potential of \TRIANGLE{}s is illustrated in~\autoref{tab:top_triangles}, where we show the links that could benefit the most from using \BESTTRIANGLE{}s at the moment the snapshot was taken.
Although these are extreme cases, our experience with the measurement system shows that this is a typical picture of the clouds' networks; at any given time, there are some links that experience painful difficulties that can be solved with \BESTTRIANGLE{}s that cut at least 50\% from the direct RTT.

Note that the links in~\autoref{tab:top_triangles} are all cross-cloud, meaning that the source and destination do not belong to the same cloud provider.
This is in accordance with the intuition that intra-cloud links have lower worst case RTTs than cross-cloud links. There are also much more inter-cloud links.
Not only that, in some cases the \BESTRELAY{} is a region that belongs to a 3$^{rd}$ cloud, different from the cloud of the source or the destination of the link.
More about this matter can be found in~\autoref{sec:cloud_level_analysis}, where we systematically analyze the relationships between the clouds. 

\begin{table*}[!h]
\centering
\begin{tabular}{ l l l c c c }
	&		&		&	Direct 	&	\BESTTRIANGLEUC{}	&	Reduced 	\\
Source	&	Destination	&	\BESTRELAYUC{} 	&	 RTT[ms]	&	 RTT[ms]	&	 RTT [\%]	\\
\hline \hline
aws.ca-central-1	&	azure.canadaeast	&	gcp.northamerica-northeast1	&	43.9	&	6.6	&	85\%	\\
azure.canadaeast	&	aws.ca-central-1	&	gcp.northamerica-northeast1	&	43	&	6.9	&	84\%	\\
gcp.asia-south1	&	aws.eu-west-3	&	aws.ap-south-1	&	353.4	&	107.3	&	69.6\%	\\
aws.eu-west-3	&	gcp.asia-south1	&	aws.ap-south-1	&	353.3	&	109	&	69.1\%	\\
azure.canadacentral	&	aws.ca-central-1	&	gcp.northamerica-northeast1	&	32	&	10	&	68.8\%	\\
gcp.asia-south1	&	aws.eu-central-1	&	aws.ap-south-1	&	351.6	&	112.8	&	67.9\%	\\
aws.eu-central-1	&	gcp.asia-south1	&	aws.ap-south-1	&	351.6	&	113	&	67.9\%	\\
aws.ca-central-1	&	azure.canadacentral	&	gcp.northamerica-northeast1	&	31.9	&	10.4	&	67.4\%	\\
gcp.asia-south1	&	aws.eu-west-2	&	aws.ap-south-1	&	339.8	&	115.4	&	66\%	\\
aws.eu-west-2	&	gcp.asia-south1	&	aws.ap-south-1	&	339.8	&	115.7	&	66\%	\\
gcp.asia-south1	&	aws.eu-west-1	&	aws.ap-south-1	&	349.9	&	123.1	&	64.8\%	\\
aws.eu-west-1	&	gcp.asia-south1	&	aws.ap-south-1	&	349.9	&	123.4	&	64.7\%	\\
azure.francecentral	&	gcp.asia-south1	&	azure.westindia	&	227.8	&	108.9	&	52.2\%	\\
gcp.asia-south1	&	azure.francecentral	&	azure.westindia	&	228.2	&	109.6	&	52\%	\\
aws.us-east-1	&	gcp.northamerica-northeast1	&	gcp.us-east4	&	30.3	&	14.6	&	51.8\%	\\
\end{tabular}
\caption{Several top \BESTTRIANGLE{}s, each saving more than 50\% of a direct RTT of a link. The \textit{Reduced} column shows the RTT reduction in percentages, relative to the \textit{Direct RTT}. Snapshot time 28-Oct-2018 09:00 UTC.}
\label{tab:top_triangles}
\end{table*}

\subsection{Characterizing Regions}\label{sec:charecterizing_regions}

After we established a procedure for finding \TRIANGLE{}s, we would like to evaluate how many times each of the clouds' regions is selected to be a \RELAY{} of a \TRIANGLE{}. 
We term a region that is serving as a \RELAY{} in a \TRIANGLE{}, and not necessarily a \BESTTRIANGLE{}, as \textit{beneficial} to other regions.

\autoref{tab:triangles_regions_helping_others} lists some of the 59 regions, showing how much help each region can provide as a \RELAY{} to links between all clouds, and how much it can help for links of each cloud.
To get a better understanding of the table's structure, we will now closely examine the first row, pertaining to the region azure.francecentral.
According to the \textit{\_any links} column, it can help 391 of the 3306 links that connect any cloud with any cloud, where both source and destination are not azure.francecentral.
By can \textit{help} we mean that it is a \RELAY{} in any \TRIANGLE{}, and not necessarily a \BESTTRIANGLE{}.
All the 391 RTT reductions, relative to each link's direct RTT, are summed up and displayed in the \textit{\_any reduced} column.
Similarly, the following columns detail how much help this region can give specific clouds, in links where both source and destination belong to the same cloud.
For example, the table shows that this region can help 17 of 210 links in AWS.

The intra-cloud results in~\autoref{tab:triangles_regions_helping_others} are a bit of a surprise.
These results indicate that many regions may help traffic of their own cloud! This phenomenon is surprising in a proprietary network, where routing is planned by a single organization to be optimal.
Even more surprising are the cross-cloud results, that show many regions helping intra-cloud traffic of clouds they do not belong to, involving two cross-cloud links that one may expect to be  slower.


\begin{table*}[!ht]
\centering
\resizebox{\textwidth}{!}{\begin{tabular}{ l | c c | c c | c c | c c }
\RELAYUC{} Region	&	\_any links	&	\_any reduced	&	aws links	&	aws reduced	&	azure links	&	azure reduced	&	gcp links	&	gcp reduced	\\
\hline \hline
azure.francecentral	&	391 of 3306	&	12.4 s	&	17 of 210	&	309.6 ms	&	65 of 650	&	386.6 ms	&	30 of 272	&	1.6 s	\\
azure.uksouth	&	355 of 3306	&	12.3 s	&	18 of 210	&	294.2 ms	&	39 of 650	&	486.0 ms	&	40 of 272	&	1.9 s	\\
azure.ukwest	&	285 of 3306	&	10.7 s	&	12 of 210	&	239.7 ms	&	32 of 650	&	316.2 ms	&	38 of 272	&	1.7 s	\\
aws.eu-west-2	&	336 of 3306	&	9.2 s	&	5 of 182	&	20.1 ms	&	39 of 702	&	431.0 ms	&	1 of 272	&	2.7 ms	\\
azure.westeurope	&	194 of 3306	&	9.1 s	&	8 of 210	&	153.6 ms	&	1 of 650	&	6.5 ms	&	34 of 272	&	1.5 s	\\
aws.ap-south-1	&	277 of 3306	&	9.1 s	&	9 of 182	&	28.6 ms	&	39 of 702	&	451.3 ms	&	24 of 272	&	1.4 s	\\
azure.westindia	&	244 of 3306	&	8.9 s	&	12 of 210	&	94.7 ms	&	41 of 650	&	428.1 ms	&	23 of 272	&	1.5 s	\\
azure.northeurope	&	181 of 3306	&	8.0 s	&	7 of 210	&	98.5 ms	&	8 of 650	&	78.9 ms	&	20 of 272	&	1.3 s	\\
azure.centralindia	&	176 of 3306	&	8.0 s	&	8 of 210	&	96.3 ms	&	22 of 650	&	259.4 ms	&	20 of 272	&	1.4 s	\\
azure.southindia	&	254 of 3306	&	6.4 s	&	9 of 210	&	97.8 ms	&	72 of 650	&	718.0 ms	&	11 of 272	&	724.2 ms	\\
azure.southeastasia	&	187 of 3306	&	6.0 s	&	15 of 210	&	144.2 ms	&	7 of 650	&	92.9 ms	&	20 of 272	&	1.0 s	\\
aws.ap-southeast-1	&	202 of 3306	&	4.9 s	&	4 of 182	&	16.8 ms	&	25 of 702	&	397.5 ms	&	- of 272	&	-	\\
aws.us-east-1	&	402 of 3306	&	4.8 s	&	15 of 182	&	79.9 ms	&	86 of 702	&	724.1 ms	&	- of 272	&	-	\\
gcp.us-east4	&	350 of 3306	&	3.7 s	&	34 of 210	&	329.6 ms	&	82 of 702	&	675.5 ms	&	2 of 240	&	3.2 ms	\\
gcp.asia-northeast1	&	359 of 3306	&	3.5 s	&	24 of 210	&	160.6 ms	&	62 of 702	&	326.8 ms	&	- of 240	&	-	\\
gcp.us-central1	&	272 of 3306	&	3.5 s	&	32 of 210	&	257.6 ms	&	11 of 702	&	74.2 ms	&	19 of 240	&	42.9 ms	\\
gcp.australia-southeast1	&	136 of 3306	&	3.4 s	&	10 of 210	&	331.2 ms	&	30 of 702	&	288.4 ms	&	- of 240	&	-	\\
azure.eastus	&	194 of 3306	&	3.2 s	&	23 of 210	&	310.1 ms	&	7 of 650	&	17.8 ms	&	30 of 272	&	626.4 ms	\\
gcp.us-west2	&	165 of 3306	&	3.1 s	&	20 of 210	&	150.3 ms	&	11 of 702	&	54.8 ms	&	3 of 240	&	5.9 ms	\\
gcp.us-west1	&	206 of 3306	&	2.9 s	&	26 of 210	&	182.6 ms	&	11 of 702	&	58.1 ms	&	10 of 240	&	29.6 ms	\\
aws.us-west-1	&	150 of 3306	&	2.8 s	&	2 of 182	&	11.1 ms	&	24 of 702	&	137.0 ms	&	- of 272	&	-	\\
gcp.asia-southeast1	&	171 of 3306	&	2.6 s	&	11 of 210	&	357.5 ms	&	32 of 702	&	187.2 ms	&	3 of 240	&	10.2 ms	\\
azure.eastasia	&	153 of 3306	&	2.6 s	&	6 of 210	&	27.8 ms	&	- of 650	&	-	&	28 of 272	&	243.3 ms	\\
\end{tabular}}
\caption{Characterizing regions - these are some of the most beneficial regions, in terms of RTT reduction in \TRIANGLE{}s (not necessarily best \TRIANGLE{}s). The \textit{$X$ links} columns show in how many $Y$ links of the $Z$ relevant links the region was beneficial as a \RELAY{} in a \TRIANGLE{}. The \textit{$X$ reduced} shows the sum of all these RTT reductions. When $X$ is either aws, azure, or gcp, the relevant links or only those inside the cloud (intra-cloud). Snapshot time 28-Oct-2018 20:00 UTC.}
\label{tab:triangles_regions_helping_others}
\end{table*}

\subsection{Cloud-Level Analysis}\label{sec:cloud_level_analysis}

\begin{figure}
    \centering
    \includegraphics[width=\columnwidth]{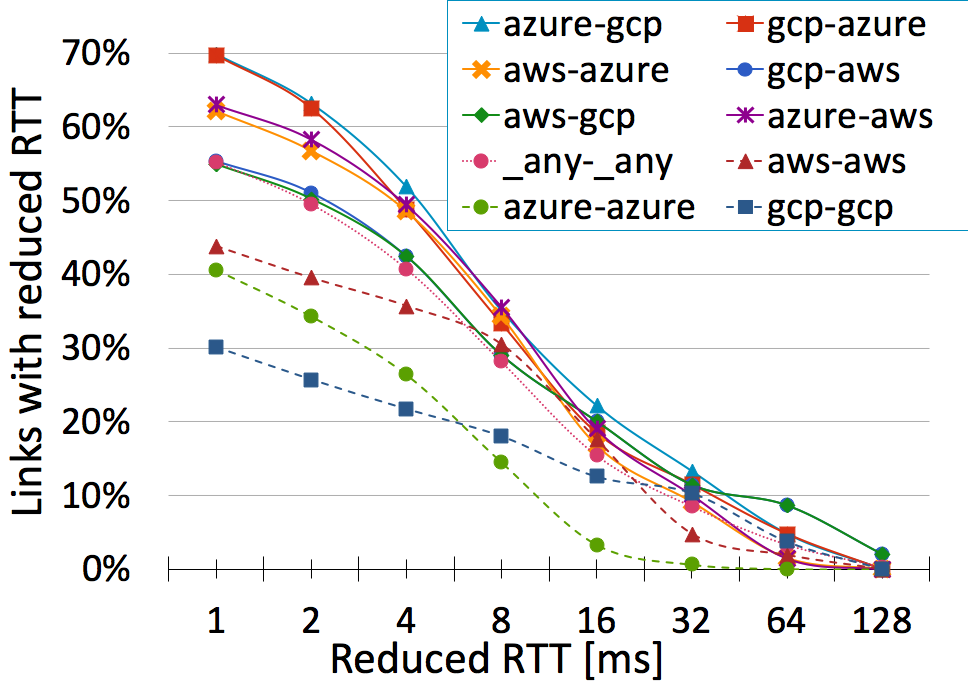}
    \vspace{-25pt}
    \caption{Cloud-level \TRIANGLE{}s CDF for all links. Dashed lines represent the intra-cloud links, while solid lines stand for cross-cloud links. The \_any-\_any line represents all links.}
    \label{fig:clouds_triangles_cdf}
\end{figure}

The surprising region-level results presented above motivated us to quantify \TRIANGLE{}s on a cloud-level basis. 
A wider look at the potential of \TRIANGLE{}s is illustrated in a cloud-level summary in \autoref{tab:clouds_triangles_summary} and \autoref{fig:clouds_triangles_cdf}.
The first row (\_any to \_any) in \autoref{tab:clouds_triangles_summary} shows that the overall percentage of links that can benefit from the usage of a \TRIANGLE{} is 54\%.
Looking at sub-categories of links, according to the cloud provider of the source and destination, we find that 29\%-45\% of the intra-cloud's links, when both the source and the destination belong to the same cloud, can still benefit from \TRIANGLE{}s.
Naturally, cross-cloud links, where source and destination belong to different cloud providers, may benefit even more; we see 53\%-70\% of the links reduced RTT thanks to \TRIANGLE{}s, depending on the specific cloud combination. \autoref{fig:clouds_triangles_cdf} depicts the percentage of links in each category that improve by at least a given amount of delay by using \TRIANGLE{}s. For example, we can see that there are around 10\% of all links (\_any to \_any) that gain at least 32 milliseconds using \TRIANGLE{}s.


\begin{table}[!h]
\centering
\begin{tabular}{ l l l c c c }
Source & Dest. & Links & Reduced RTT & Links benefit \\
\hline \hline
\_any  & \_any &  3422 &       7.3\% & 54\% \\
aws    & aws   &   210 &       4.8\% & 45\% \\
aws    & azure &   405 &       8.6\% & 60\% \\
aws    & gcp   &   255 &      11.2\% & 53\% \\
azure  & aws   &   405 &       9.3\% & 63\% \\
azure  & azure &   702 &       2.4\% & 40\% \\
azure  & gcp   &   459 &       8.6\% & 70\% \\
gcp    & aws   &   255 &      11.2\% & 53\% \\
gcp    & azure &   459 &       8.3\% & 67\% \\
gcp    & gcp   &   272 &       5.6\% & 29\% \\
\end{tabular}
\caption{Cloud-level \TRIANGLE{}s summary, organized by source and destination clouds. The \textit{Links} column says how many links connect the \textit{Source} and \textit{Destination} clouds. The \textit{Links benefit} column says how many of the \textit{Links} can benefit from \TRIANGLE{}s. The \textit{Reduced RTT} is the overall RTT reduction when using \TRIANGLE{}s, relative to the direct RTT in all \textit{Links}.}
\label{tab:clouds_triangles_summary}
\end{table}


\IC{(1) remove the "log scale bucket" from the X axis, just leave Reduced RTT (ms) (2) in the cation, what you mean by "Cloud-level triangles CDF for any relay", does not seem like a CDF graph and it's not per relay. It seems to be per link.}

\subsection{\RELAYUC{}s Locations}

\begin{figure}[!h]
    \centering
    \includegraphics[width=\columnwidth]{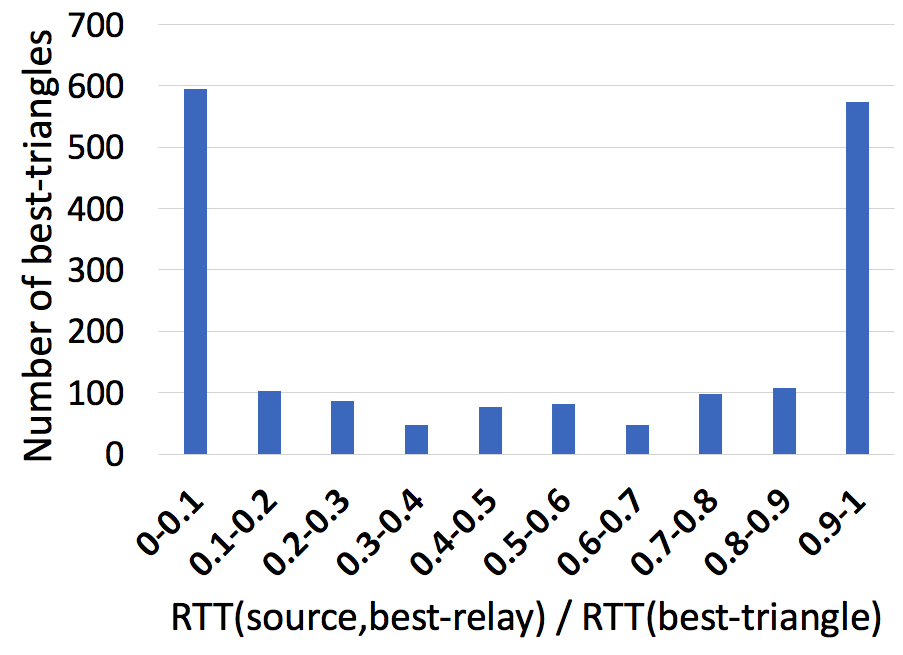}
    \vspace{-25pt}
    \caption{Histogram of the \BESTTRIANGLE{}s vs. the relative RTT of the \BESTRELAY{} to the source region out of the total RTT of the \BESTTRIANGLE{}. It shows that in almost all the \BESTTRIANGLE{}s, the \BESTRELAY{} is either close to the source (up to 0.1) or to the destination (more than 0.9), so it is usually not geographically located in between the two (0.1 to 0.9).}
    \label{fig:triangles_relay_location}
\end{figure}

\YBI{I think it is. Preferably, if figure 10 will be placed right after the paragraph.}
Interestingly, in most \BESTTRIANGLE{}s we have found, the \RELAY{} is located in a region that is geographically close to either the source or the destination regions.
\autoref{fig:triangles_relay_location} illustrates the distance between a \RELAY{} and the source region, in terms of RTT of \BESTTRIANGLE{}s.
The ratio simply shows how much of the \BESTTRIANGLE{}'s overall RTT is related to the first segment, which is the source-to-\RELAY{} part.

\subsection{Considering More \RELAYUC{}s}

Since adding a \RELAY{} seems like a good idea in many cases, one may ask if adding more \RELAY{}s pays-off.
Intuitively, it stands to reason that adding a 2$^{nd}$ \RELAY{} would not be as beneficial as the first one that established the \TRIANGLE{}.
To examine that assumption, we run a shortest path algorithm with 2 and 3 \RELAY{}s.
The results are presented in~\autoref{tab:recursive_triangles}.
As the table shows, adding a 2$^{nd}$ \RELAY{} can further reduce the RTT for about 10\% of the links, reducing their RTT by less than 1\%.
Although it seems like there is no case for such complication, we point out that in some cases the option to add more \RELAY{}s may be crucial: when communication interruptions affect several regions at once, a complex long detour might be the only way to go around a traffic bottleneck, even if it only lasts for a few minutes. 

\begin{table}
\begin{center}
\resizebox{\columnwidth}{!}{
\begin{tabular}{ l c c c }

\textbf{Relays} & \textbf{Additional RTT reduction} & \textbf{Reduced links} \\
                & \textbf{relative to direct}   & \\
\hline \hline
0 (direct)       &         & 3422 \\
1 (\TRIANGLE{}s) & 6.544\% & 1816 \\
2                & 0.688\% & 363 \\
3                & 0.008\% &  16 \\   
\end{tabular}
}
\caption{The diminishing returns of more \RELAY{}s.}
\label{tab:recursive_triangles}
\end{center}
\end{table}
\section{Related work}


Internal data center traffic has been studies extensively over the past few years~\cite{kandula2009nature, benson2009understanding, benson2010network}. 
Cross-cloud traffic, however, received much less attention from the research community.
Yahoo! analyzed datasets~\cite{chen2011first} of inbound and outbound traffic collected at border routers between five Yahoo! data centers located worldwide to try and characterize inter data center traffic. They differentiate between client-triggered and background traffic related to internal tasks such as backups. 

A recent study~\cite{dhamdhere2018inferring} focuses on persistent inter-domain links between different ISPs. The methodology they used for measuring congestion, namely \textit{Time Sequence Latency Probes} (TSLP)~\cite{dhamdheremeasuring}, assumes that a ping to the near and far routers travels on the same path. It is a valid assumption when measuring congestion on the Internet. In more modern settings, on the other hand, especially in clouds, this assumption no longer holds. 

CloudHarmony~\cite{cloudharmoney} maintains a network performance measurement portal, which provides an aggregated latency statistics for a predefined duration. CloudHarmony performs its testing using the RIPE Atlas Internet measurement network, which consists of approximately 16284 public test probes. The tests include CPU, disk IO, memory, and other performance factors. CloudHarmony's network tests differ from our work in several aspects. First, CloudHarmony provides the raw measurements only, without any additional evaluation methods or procedures. In addition, CloudHarmony tests the connection of a client to different regions in various clouds, but it does not include cross-cloud or intra-cloud measurements. Finally, while CloudCast collects measurements automatically on a per-minute basis, CloudHarmony performed tests only on demand.

Broader aspects of large data center and cloud optimization have been pushed to the forefront ever since the appearance of public clouds, captivating both the research community and the industry. Challenges such as auto-scaling~\cite{mao2011auto,mukherjee2017optimal}, job scheduling~\cite{shifrin2013}, energy consumption~\cite{zhang2011greenware} to name a few are interleaved with budgeting issues~\cite{greenberg2008cost} and other concerns.
When it comes to service pricing, as mentioned previously, cloud providers do not offer at this time quality of service guarantees in terms of data delivery. This could be due to the highly diverse nature of cloud traffic and the resulting conflicting demands introduce when defining SLAs~\cite{maltz2013challenges}.

\section{conclusion}\label{sec:conclusion}

In this work, we introduced CloudCast, a measurements and analysis system, spanning all available regions of the three leading cloud providers. The system monitors a total of 3384 links, collecting about 2GB of data every month.
Our system performs both real time and offline evaluation of the measurements collected.

Through careful analysis of the data collected by CloudCast, we were able to diagnose several trends, which we believe represent the state of cloud networks today. We offer new methodologies for the examination and fair comparison of cloud networking.  Finally, we utilize one of our new methodologies, namely \TRIANGLE{}s, to offer a possible solution for overcoming temporary shortages in numerous regions.

Considering the sheer amount and the extent of the data we collected and will collect in the future, we believe it can fuel a multitude of follow-up works. We plan to extend our measurement system across additional public clouds and SaaS. 


\bibliographystyle{plain}
\bibliography{main}

\end{document}